\documentclass[aps,twocolumn,groupedaddress,showpacs,amssymb,prb,floatfix,preprintnumbers]{revtex4}
\usepackage{graphicx,color}
\usepackage{amssymb}

\newcommand{\slrr}      {$T_1^{-1}$}

\newcommand{\tn}     {$T_{\rm N}$}

\newcommand{\slrrtext}  {spin-lattice-relaxation rate}

\newcommand{\lanio}     {La$_{4}$Ni$_3$O$_8$}

\begin{document}

\thispagestyle{myheadings}

\title{NMR evidence for antiferromagnetic order in a planar 3d$^9$/3d$^8$ nickelate}

\author
{N. ApRoberts-Warren$^1$, A. Dioguardi$^1$, V. V. Poltavets$^2$, M. Greenblatt$^3$, and N. J. Curro$^{1}$ \email{curro@physics.ucdavis.edu}}

\affiliation{$^{1}$Department of Physics, University of California, Davis, CA 95616, USA\\
 $^{2}$Department of Chemistry, Michigan State University, East Lansing, MI 48824, USA\\
$^{3}$Department of Chemistry and Chemical Biology, Rutgers, The State University of New Jersey, 610 Taylor Road, Piscataway, NJ 08854, USA}

\date{\today}

\begin{abstract}
We report $^{139}$La Nuclear Magnetic Resonance (NMR) data in La$_{4}$Ni$_3$O$_8$ which reveals the presence of antiferromagnetic order below $T_N\sim 105$ K.  This compound contains two-dimensional layers of NiO$_2$ that are isostructural to the copper oxide planes of the high temperature superconductors.  This compound is remarkable because the average Ni valence of $1.33+$ for formally 2Ni$^{1+}$ + Ni$^{2+}$ indicates a 3d$^9$/3d$^8$ electronic configuration. Nickel oxides with Ni$^{1+}$ valence are rare and unstable, yet may provide new routes to high temperature superconductivity.  La$_{4}$Ni$_3$O$_8$, with  antiferromagnetic order of $S=1/2$ Ni$^{1+}$ spins, is analogous to the parent compounds of the cuprate superconductors.  Our data clearly reveal dramatic spectral changes and low energy antiferromagnetic correlations associated with the onset of long range order below $T_N$.

\end{abstract}

\pacs{ 76.60.-k, 71.27.+a, 75.50.Ee, 75.47.Lx}

\maketitle

Prior to their seminal discovery of high temperature superconductivity, Bednorz and M\"{u}ller initially focused their search for superconductivity on the La-Ni-O system with Ni in the $2+/3+$ valence. \cite{BednorzNobelLecture} Despite intensive efforts to find other transition metal analogs to the cuprates, the only other systems known to exhibit superconductivity are the  cobaltates and the iron oxypnictides.  On the other hand, the nickelates have the greatest potential to exhibit physics similar to the cuprates. Theoretical studies show that only nickelates with Ni$^{1+}$ (3d$^9$, $S = 1/2$) in a square planar coordination can form an antiferromagnetic (AFM) insulator directly analogous to the parent (undoped) cuprates. \cite{anisimov} A square planar coordination is required to force Ni$^{2+}$(3d$^8$) into the low spin ($S = 0$) configuration, while in octahedral Cu$^{3+}$(3d$^{8}$) oxides, the higher oxidation state and consequently larger crystal field splitting result in a low spin state. Nickelates with Ni$^{1+}$/Ni$^{2+}$ are metastable, and there are only a few compounds with infinite NiO$_2$ planes known to exist. \cite{PoltavetsLa3Ni2O6} Here we report NMR data that reveal antiferromagnetic order below \tn$=105$ K in \lanio, a two dimensional square planar system with Ni$^{1+/2+}$. \cite{poltavets2010}

\begin{figure}
\includegraphics[width=1.0\linewidth]{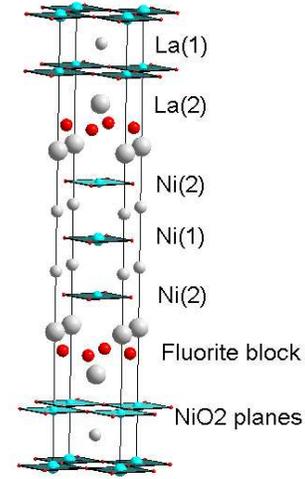}
\caption{\label{fig:structure} (color online) The unit cell of \lanio, showing the three sets of NiO$_2$ planes separated by LaO fluorite blocks.  There are two La sites, with La(1) located between the NiO$_2$ planes, and La(2) located in the LaO blocks. }
\vspace{-0.15in}
\end{figure}

Polycrystalline samples of \lanio\ (tetragonal with space group $I4/mmm$) were prepared by low temperature reduction of the Ruddlesden-Popper  La$_4$Ni$_3$O$_{10}$ $n=3$ phase as described in \cite{poltavets2007}.  \lanio, shown in Fig. \ref{fig:structure}, exhibits infinite layers of NiO$_2$ separated by LaO fluorite blocks, and contains both an inner and two outer NiO$_2$ planes.  There are three Ni atoms per formula unit, with an average valence of $+1.33$. It is unclear how these charges are distributed, but a natural physical picture is that the 3d$_{x^2-y^2}$ orbitals of Ni$^{1+}$ ions hybridize with the O 2p orbitals, and the resulting system is hole doped with 0.33 holes per unit cell.  In a localized picture the holes may reside entirely on the Ni(1) in the inner plane to create Ni$^{2+}$ (3d$^8$), leaving the two outer planes with Ni$^{1+}$ (3d$^9$, $S=1/2$).  For  Ni$^{2+}$ with 3d$^8$ Hund's coupling generally drives an $S=1$ configuration; however for planar coordination in the absence of apical oxygens the large crystal field splitting between the $x^2-y^2$ and the $z^2$ d-orbitals forces Ni$^{2+}$ into the $S=0$ state. \cite{anisimov} On the other hand, in an itinerant picture the excess charge may be shared among three Ni bands, and the magnetic behavior will be driven by band structure effects.

Both scenarios predict a spin degree of freedom for some of the Ni ions, and this prediction is borne out by specific heat and magnetization measurements which reveal a second order phase transition at 105 K. \cite{poltavets2010} The magnetic entropy loss associated with this transition is consistent with both the localized picture with roughly one-half of $R\ln 2$ for each of the two 3d$^9$ Ni(2) sites or roughly one-third of $R\ln 2$ per each Ni band in an itinerant picture.  The magnetization is suppressed by 15\% below \tn.  Resistivity data reveal insulating behavior with a change of slope at \tn. \cite{poltavets2010} However, the granular nature of the polycrystalline sample may have precluded intrinsic measurements of the resistivity.  Indeed, our measurements of the \slrrtext\ suggest metallic behavior, as discussed  below.
\begin{figure}
\includegraphics[width=1.0\linewidth]{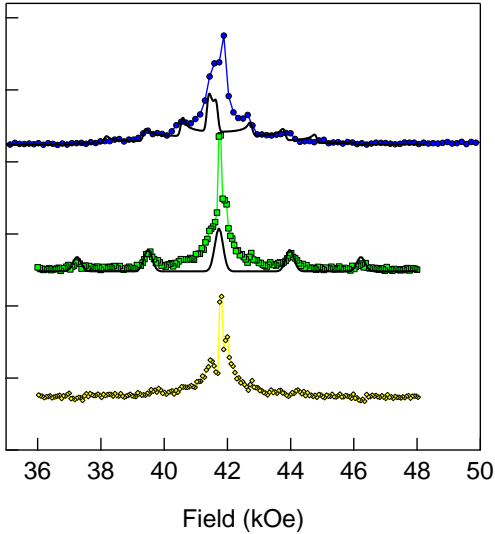}
\caption{\label{fig:powdercomparison} (color online) Field-swept spectra at 25.1 MHz fixed frequency in a random powder (upper spectrum, blue) and in an aligned powder (middle spectrum, green).  The solid black lines are fits to the La(1) spectrum in each case. The bottom spectrum (yellow) is the difference between the aligned spectrum and the fit to the La(1) contribution, and should represent the spectrum of the La(2).}
\vspace{-0.15in}
\end{figure}

In order to investigate the nature of the phase transition in more detail we have carried out $^{139}$La NMR ($I=7/2$) as a function of temperature.  Approximately 100mg of powder were mixed with Stycast epoxy and cured in an external field in order to align the powder and improve the spectra resolution.  Fig. \ref{fig:powdercomparison} shows the La spectra in both the random powder and the aligned samples. Spectra were obtained by integrating the spin echo intensity as a function of applied magnetic field, $H_0$, at fixed frequency. There are clear differences, with well resolved satellites in the aligned spectrum.  These spectra indicate successful alignment and suggest that the magnetic susceptibility is highly anisotropic, with $\chi_c/\chi_{ab}\sim 5-10$.  The aligned spectrum reveals two distinct La sites, consistent with the structure, with  La(1) located between the NiO$_2$ planes and La(2) located in the LaO fluorite blocks (Fig. \ref{fig:structure}).  Both sites have axial symmetry and the La nuclei are sensitive to the local electric field gradient (EFG). The resonances are determined from the nuclear Hamiltonian:  $\mathcal{H} = \gamma\hbar\mathbf{H}_0\cdot(1 + \mathbf{K})\cdot\mathbf{\hat{I}} + (h\nu_Q/6)(3\hat{I}_c^2 - \hat{I}^2)$,
where $\gamma$ is the gyromagnetic ratio, $K$ is the Knight shift, $\nu_Q= 3eQV_{cc}/2I(2I-1)h$ is the quadrupolar frequency, $e$ is the electron charge, $Q$ is the quadrupolar moment, and $V_{cc}$ is the component of the EFG tensor corresponding to the tetragonal $c$ axis of the unit cell. \cite{CPSbook} For an applied field along the $c$ direction and fixed frequency $f$, the resonance fields for each site are given by: $H_n = f/\gamma(1+K) + n\nu_Q/\gamma$, where $n=-3,\cdots+3$.  The solid lines in Fig. \ref{fig:powdercomparison} are fits with $\nu_Q = 1.35(5)$ MHz, tentatively labeled as the A site. The lower spectrum (yellow) shows the difference between the data and the fit and correspond to the other La site (the B site).  The satellites are poorly resolved, but we estimate  $\nu_Q\sim 20(10)$ kHz. Calculations of the EFG using Wien2K code in the paramagnetic state give $V_{cc}(1) = -3.33\times 10^{21}$ V/m$^2$ ($\nu_Q(1) = 0.5$ MHz) and
$V_{cc}(2) = -0.15\times 10^{21}$ V/m$^2$ ($\nu_Q(2) = 20$ kHz) for the La(1) and La(2) sites, respectively.  We thus assign the A site to La(1) and the B site to La(2).

\begin{figure}
\includegraphics[width=1.0\linewidth]{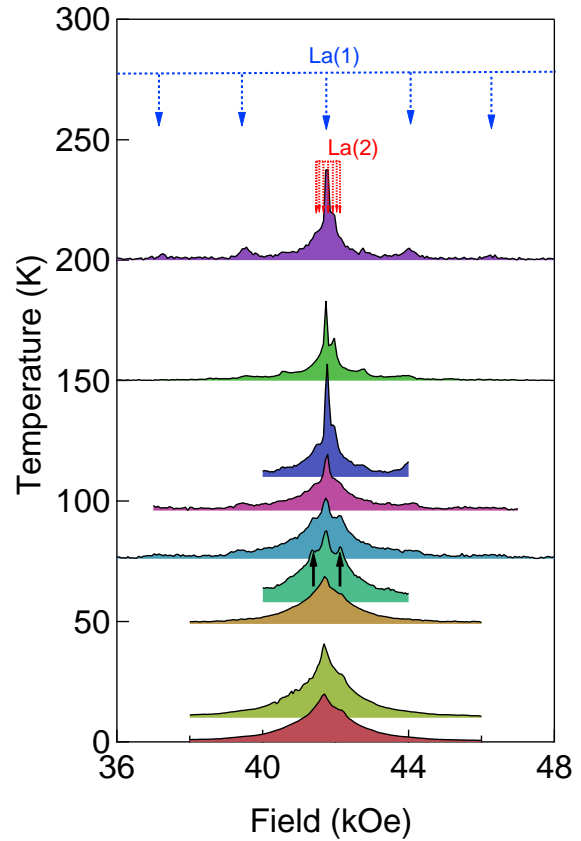}
\caption{\label{fig:spectra} (color online) Field-swept spectra at 25.1 MHz fixed frequency in an aligned powder of \lanio.  The positions of the various satellite transitions for each site are indicated, and the solid black arrows at the 60K spectrum indicate the positions of the peaks plotted in Fig. \ref{fig:orderparameter}.}
\vspace{-0.15in}
\end{figure}

\begin{figure}
\includegraphics[width=\linewidth]{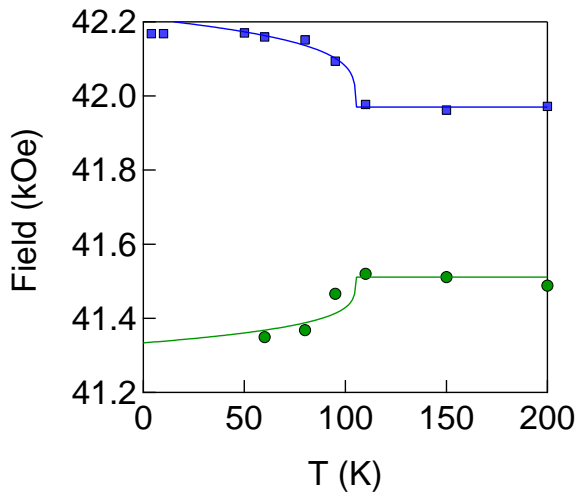}
\caption{\label{fig:orderparameter} (color online) Positions of the upper and lower sub-peaks in the La spectrum, as indicated by the arrows in Fig. \ref{fig:spectra}. The solid lines are fits as described in the text. }
\vspace{-0.15in}
\end{figure}

Figure \ref{fig:spectra} shows the full La spectrum in the aligned sample as a function of temperature.  Below the phase transition the La(1) spectrum is washed out by the presence of a broad distribution of static hyperfine fields, whereas the La(2) resonance remains visible. The powder nature of the sample coupled with the small value of the La(2) EFG renders the La(2) spectrum nearly featureless, but we can identify two sub-peaks in the spectrum which develop a significant temperature dependence below 105 K (see arrows in Fig. \ref{fig:spectra}). We find an internal field on the order of 200 Oe develops at the La(2) site  (see Fig. \ref{fig:orderparameter}).  This internal field arises from the non-cancelation of the transferred hyperfine fields at the La(2) site from ordered Ni moments, and is thus a direct measure of the order parameter of this system. Similar behavior was observed in powder spectra of LaFeAsO below \tn. \cite{Nakai2008} The order develops with a second-order temperature dependence that is well fit by the expression $M(T)/M_0=(1-(T/T_N)^{\beta}$ with $\beta = 0.25$.  This value of $\beta$ is less than the mean field value of 1/2, and may be related to the two-dimensional nature of the NiO$_2$ planes. A similar field develops at the La(1) site, but since the hyperfine fields are much larger in the La(1) case the spectrum is broadened considerably. It is difficult to extract information about the magnetic structure in this case, but the broad nature of the La(1) spectrum suggest an incommensurate wavevector. Indeed, density functional (DFT) calculations indicate a spin density wave (SDW) with ordering wavevector $\mathbf{Q} = (\pi/3a, \pi/3a, 0)$, although direct neutron scattering measurements show no ordering at this particular wavevector. \cite{poltavets2010}

In order to understand the magnetic order and the antiferromagnetic fluctuations it is important to consider the hyperfine coupling to the two La sites.  To lowest order we can assume isotropic transferred hyperfine coupling to each of the nearest neighbor Ni sites. La(1) is located symmetrically between 4 nearest neighbor Ni(1) spins (inner plane) and 4 nearest neighbor Ni(2) spins (outer plane).  La(2) is located symmetrically between four nearest neighbor Ni(2) spins. For staggered antiferromagnetic order, $\mathbf{Q}=(\pi/a,\pi/a)$, as in the parent compounds of the high-$T_c$ cuprates, the hyperfine field  will vanish by symmetry at both La sites.  However, for $\mathbf{Q} = (\pi/3a, \pi/3a, 0)$ ordering, there are multiple La(1) and La(2) sites with different hyperfine fields.  This distribution of hyperfine fields gives rise to the broad spectra observed in Fig. \ref{fig:spectra}, and fluctuations of these fields in the paramagnetic state will contribute to the \slrrtext.

\begin{figure}
\includegraphics[width=0.9\linewidth]{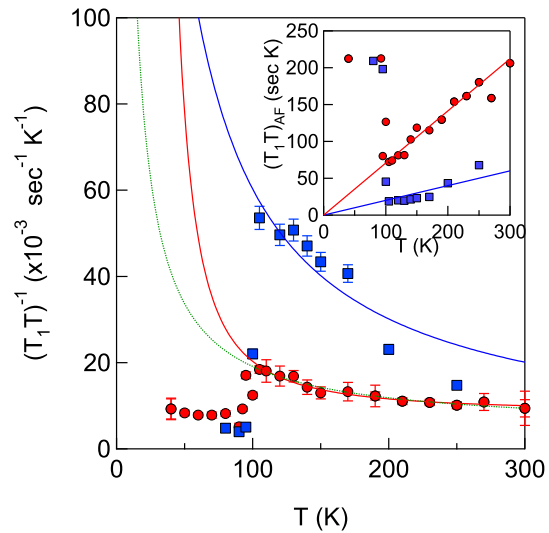}
\caption{\label{fig:T1} (color online) $(T_1T)^{-1}$ versus temperature at the La(1) site (blue $\blacksquare$) and La(2) site (red $\bullet$). Solid lines are fits as described in the text, with $a=0$, $b=6.0(1)$ sec$^{-1}$ and $T_0=0$ for La(1) and  $a=0.0045(4)$ sec$^{-1}$ K$^{-1}$, $b=1.46(7)$ sec$^{-1}$ and $T_0=0$ for La(2). THe dotted line is a fit to the 2D Heisenberg expression as discussed in the text with $J=129(5)$ K. INSET: $(T_1T)_{\rm AF} = T_1T - a^{-1}$ versus temperature. Solid lines are guides to the eye.}
\vspace{-0.15in}
\end{figure}

In order to characterize the dynamics of the phase transition we have measured the \slrrtext, \slrr, at both sites (see Fig. \ref{fig:T1}).  \slrr\ was measured by inversion recovery, and fitting the magnetization to the expression to the standard expression for $I=7/2$. For the La(1) site \slrr\ was measured at the first satellite transition ($-3/2\leftrightarrow-1/2$) at 44kOe (see Fig. \ref{fig:spectra}), and for the La(2) \slrr\ was measured at the central peak.  Because of spectral overlap at the La(2) it is likely that several of the La(2) satellites as well as part of the La(1) central transition were also inverted and thus the \slrr\ data is not purely the La(2).  Nevertheless both sites exhibit strong temperature dependences, and dramatic reductions at \tn, which clearly establishes this phase transition as intrinsic.

The strong temperature dependence above \tn\ reveals the presence of critical fluctuations up to 300 K.   \slrr\ is given by:
\begin{equation}
\frac{1}{T_1T} = \gamma^2k_B\lim_{\omega\rightarrow 0}\sum_{\mathbf{q},\beta}F^2_{\beta}(\mathbf{q})\frac{\chi_{\beta}''(\mathbf{q},\omega)}{\hbar\omega},
\end{equation}
where the form factor $F(\mathbf{q}$) is the Fourier transform of the hyperfine coupling and $\chi_{\beta}''(\mathbf{q},\omega)$ is the dynamical susceptibility \cite{Moriya1974}.
The enhanced value of \slrr\ at the La(1) site suggests a larger form factor than at the La(2), consistent with the fact that La(1) is closer\cite{poltavets2007} and hence more strongly coupled to the Ni spin fluctuations.  Surprisingly, $(T_1T)^{-1}$ does not appear to diverge at $T_N$, as expected for a second order phase transition in 3D.  The solid line in Fig. \ref{fig:T1} is a fit to $(T_1T)^{-1} = a + b/(T-T_0)$ above \tn, where the first term represents a Korringa contribution from spin-spin scattering with quasiparticles at the Fermi surface, and the second term represents the contribution from 2D antiferromagnetic spin fluctuations. \cite{Moriya1974,MMPT1inYBCO}  We find that $T_0=0$ fits the data best, suggesting that the antiferromagnetic spin fluctuations are truly 2D, and would not be expected to exhibit long range order at finite $T$. \cite{CHN}  The dotted line shows a fit to $(T_1T)^{-1} = a + b\sqrt{x}e^{1/x}/(1+x)^3$ with $x = T/1.13J$, as expected for a 2D Heisenberg antiferromagnet with $S=1/2$ with $J=129(5)$ K. \cite{ChakOrbach}  Both expressions fit the data well, and suggest that the
spin fluctuations are analogous to the high temperature superconducting cuprates. In the cuprates antiferromagnetic correlations of Cu spins in 2D CuO$_2$ planes are present up to high temperatures ($J\sim 1500$ K), and long range antiferromagnetic ordering is driven by a small interplanar exchange $J_{\perp}\sim 10^{-5}J$ at $T_N\sim 300$K.  In the case of \lanio\ our data are consistent with $J \sim 130$ K, and the layered structure could give rise to strong anisotropy in the antiferromagnetic exchange coupling. Indeed, the separation between NiO$_2$ planes in the tri-layers (across the fluorite blocks) is approximately twice the separation between NiO$_2$ planes within the tri-layers.  Furthermore, the tri-layers are shifted by $(a/2,a/2)$, frustrating the antiferromagnetic interactions along the $c$ direction.

 Thus a plausible scenario might be that Ni(2) spins with $S=1/2$ are coupled within plane and weakly coupled between planes, whereas the Ni(1) site remains inactive with $S=0$.  However, the large entropy released at the ordering temperature is inconsistent with a picture of strongly anisotropic 2D spin fluctuations, in which only a fraction of $R\log 2$ is released at \tn. \cite{PinakiRajiv}  Alternative explanations may be (i) the transition is first-order and hence there are no divergent critical fluctuations, or (ii) the entropy is not associated with the Ni spins but rather is associated with the orbitals occupied by the 3d$^{8}$ Ni(1) inner planes.  Case (i) can be ruled out because low temperature neutron diffraction reveals no structural changes and the behavior of the internal field is continuous (Fig. \ref{fig:orderparameter}). \cite{poltavets2010}  For case (ii), the entropy could be associated with nearly degenerate orbitals of the Ni(1) electrons that become ordered along with the Ni(2) spins in a cooperative fashion at \tn.
   This scenario would require, however, that the $S=1$ state of the Ni(2) lie close in energy to the $S=0$ state, and there  is no evidence of $S=1$ states in the magnetic susceptibility.

  Yet another possibility is that the doped holes are itinerant, and there is little difference between the Ni(1) and Ni(2) electronic configurations.
As seen in Fig. \ref{fig:T1}, $(T_1T)^{-1}$ is approximately constant both at high temperature and at low temperature $T\ll T_N$.  This result indicates that \lanio\ is indeed a Fermi liquid for sufficiently high temperatures and that only a portion of the density of states is gapped by the SDW order.  This observation is at odds with the insulating behavior of the resistivity; however due to the metastability of \lanio, (the sample decomposes at $T > 350^{\circ}$ C), transport measurements can only be performed on polycrystalline pressed pellets where poor inter-grain electrical contacts are expected.  In fact density functional theory calculations indicate that \lanio\ is metallic, with three Fermi surfaces, one of which is strongly nested. \cite{poltavets2010}  Thus it is likely that two Fermi surfaces remain ungapped below \tn\ and contribute to Korringa relaxation at the two La sites.  On the other hand, for a second order SDW transition critical fluctuations should dominate \slrr\ above \tn, in contrast to our observations.   We expect, therefore, that the true physical picture may represent a more complex superposition of both local and itinerant character of holes.

In conclusion, we have measured La NMR at both La sites in \lanio\ and found the presence of long range antiferromagnetic order.  \slrr\ measurements strongly suggest 2D antiferromagnetic fluctuations of Ni moments, with long range order developing due to the presence of a small interplanar coupling.  This 2D layered system of $S=1/2$ Ni spins is therefore analogous to the parent antiferromagnetic state of the high temperature superconductors and potentially represents a novel route to Ni based superconductivity upon doping.  There are important differences however, as the parent cuprate is a half-filled Mott insulator, whereas \lanio\ is hole doped and possesses a Fermi surface.  Studies directed at probing the differences between the inner and outer plane Ni valences will  undoubtedly shed important new light on the magnetism of this and other transition metal oxide systems.


We thank K. Lokshin, G. Kotliar and R. Singh for stimulating discussions.


\end{document}